\documentclass[preprint,subfigure,authoryear,12pt]{elsarticle}
\usepackage{amssymb}
\journal{New Astronomy}

\usepackage[pdfborder={0 0 0 },urlcolor=blue]{hyperref}
\ifx \doiurl \undefined \def \doiurl#1{\href{http://dx.doi.org/#1}{\url{#1}}}\fi
\ifx \adsurl \undefined \def \adsurl#1{\href{http://adsabs.harvard.edu/abs/#1}{\url{#1}}}\fi

\newcommand{\etal}{{\it et al.}}

\begin{document}

\begin{frontmatter}
\title{Long-term variations in the north-south asymmetry of solar activity 
and solar cycle prediction, III: prediction for the amplitude 
of solar cycle~25}

\author{J.\ Javaraiah}

\address{Indian Institute of Astrophysics, Bengaluru - 560034, India \\
Tel: +91 80 25530672,  Fax: +91 80 25534043}

\ead{jj@iiap.res.in}

\begin{abstract}
 The  combined  Greenwich and Solar
 Optical Observing Network (SOON) sunspot group data during 
1874\,--\,2013 are analysed and studied the relatively long-term  
variations in the annual sums
  of the areas of sunspot groups in  
 $0^\circ -10^\circ$, $10^\circ - 20^\circ$, 
and $20^\circ -30^\circ$ latitude intervals  of the Sun's  northern 
and southern hemispheres. 
The variations in  the corresponding north-south differences are also studied. 
Long periodicities in these parameters are determined from the fast
 Fourier transform (FFT), maximum entropy method (MEM), and Morlet wavelet
 analysis. It is found that 
in  the difference between the sums of the  areas of the sunspot groups in 
 $0^\circ-10^\circ$ latitude intervals of northern and southern hemispheres, 
  there exist $\approx$ 9-year  periodicity  during the high activity 
 period 1940\,--\,1980 and  
 $\approx$ 12-year periodicity  
during the low activity period 1890\,--\,1939.
 It is also found that there exists a  
 high correlation (85\% from 128 data points) between the sum of the
 areas of the sunspot groups in $0^\circ - 10^\circ$ latitude interval of  the
southern hemisphere during  a $Q$th year (middle year of 3-year smoothed 
time series)  and the  annual mean    
 International Sunspot Number ($R_{\rm Z}$)
 of ($Q + 9$)th year. 
  Implication of these results
  is discussed   in the context of solar activity prediction and 
 predicted $50\pm10$ for the amplitude of solar cycle~25, 
which is about
31\% lower than the amplitude of cycle~24.
\end{abstract}


\end{frontmatter}

\parindent=0.5 cm

\newpage
\section{Introduction}
Solar activity varies on many timescales 
 \citep[from decades to stellar evolutionary timescales,][]{roz01,hath10,ay12}.
 Fig.~1 shows the long-term variations in 
 the annual mean monthly Z\"urich or International
 Sunspot Number
 ($R_{\rm Z}$)
 taken from the website,
{\tt ftp://ftp.ngdc\break .noaa.gov/STP/space-weather/solar-indices/sunspot-numbers}.
  The study of variations in the solar activity is important for understanding 
the basic mechanism of solar cycle and for predicting the level of solar activity. 

It is well believed that interactions of solar convection, magnetic field, 
 rotational and meridional flows  responsible for solar activity and 
solar cycle.
It is known that solar activity, rotation rate,  and meridional velocity are
 latitude and time dependent. Studies on latitude and time
 dependent variations of these phenomena are important for understanding the
 mechanism behind  solar cycle, which is not yet fully understood. The
 latitude and time dependencies in  these
 large-scale flows may cause   
the magnetic fields at different heliographic latitudes during
different time-intervals of  a solar cycle for contributing (/relating)  the
 activity at the same or different heliographic latitudes during its 
following cycle(s).  These different latitude bands of activity could
 produce correlations useful in predictions. 
 Recently, 
 by using the sunspot group data during the period 1874\,--\,2006, we have
 found the following results \citep[][hereafter Paper~I, Paper~II]{jj07,jj08}. 
\begin{enumerate} 
\item REL1--The sum of the areas of the sunspot groups  in
 $0^\circ$\,--\,$10^\circ$ latitude
 interval of  the Sun's  northern hemisphere and in the time-interval
of $-1.35$ year to $+2.15$ year from the time of the preceding minimum
 of a solar cycle $n$ correlate well (correlation  coefficient $r = 0.947$)
 with the  amplitude
(largest 13-month running smoothed sunspot number) of
the next cycle $n+1$. The correlation  between the 
north-south difference of the corresponding  area sums in these  
 latitude  and time intervals 
and the amplitude of cycle~$n+1$ is found to be even  much
 higher ($r = 0.968$).
 \item REL2--The sum of the areas of the sunspot groups  in
$0^\circ$\,--\,$10^\circ$ latitude
 interval of the
 southern hemisphere and in the time-interval of
 1.0 year to 1.75 year just after the time of the maximum of the cycle $n$
 correlate very well
 ($r = 0.966$)
with the amplitude of cycle $n+1$.
\end{enumerate}

The following are  the    abbreviations and their corresponding meanings
  that were  used in Papers~I and II, which are 
 also used (or mentioned) in the present paper.
\begin{itemize}
\item $n$ - the Waldmeier solar cycle number,
\item $T_{\rm m}$ - the  preceding  minimum epoch of a solar cycle,
\item $T_{\rm M}$ - the maximum epoch of a solar cycle,
\item $R_{\rm m}$ - the value of 13-month smoothed $R_{\rm Z}$ at $T_{\rm m}$,
\item $R_{\rm M}$ - the value of 13-month smoothed $R_{\rm Z}$ at $T_{\rm M}$,
\item $T_{\rm m}^*$ - the time-interval of $-1.35$ year to $+2.15$ year from $T_{\rm m}$,
\item $T_{\rm M}^*$ - the time-interval of $1.0$ year to $1.75$ year just after $T_{\rm M}$,
\item $A_{{\rm N}, n} (T_{\rm m}^*)$ - the sum of the areas of the 
sunspot groups in
 $0^\circ - 10^\circ$ latitude interval of the northern hemisphere
 during $T_{\rm m}^*$ of a cycle $n$,
\item $A_{{\rm S}, n} (T_{\rm m}^*)$ - the sum of the areas of the 
sunspot groups in
 $0^\circ - 10^\circ$ latitude interval of the southern  hemisphere
during $T_{\rm m}^*$ of a cycle $n$,
\item $A_{{\rm N}, n} (T_{\rm M}^*)$ - the sum of the areas of the 
sunspot groups in
$0^\circ - 10^\circ$ latitude interval of the northern hemisphere
during $T_{\rm M}^*$ of a cycle $n$,
\item $A_{{\rm 
 S}, n} (T_{\rm M}^*)$ - the sum of the areas of the sunspot groups in
$0^\circ - 10^\circ$ latitude interval of the southern hemisphere
during $T_{\rm M}^*$ of a cycle $n$, 
\item $\delta A_n (T^*_{\rm m})$ -  the difference, $A_{{\rm N}, n} (T_{\rm m}^*) - A_{{\rm
 S}, n} (T_{\rm m}^*)$, and 
\item $\delta A_n (T^*_{\rm M})$ -  the difference, $A_{{\rm N}, n} (T_{\rm M}^*) - A_{{\rm
 S}, n} (T_{\rm M}^*)$.
\end{itemize}

The following  equations correspond to REL1,   obtained in 
Papers~I and II (there are minor/negligible changes in the values 
of coefficients due to some minor data corrections):

\begin{equation}
R_{\rm M, n+1} = (1.7 \pm 0.2) A_{{\rm N},n} (T^*_{\rm m}) + (74 \pm 7) ,
\end{equation}

\noindent and
\begin{equation}
R_{\rm M, n+1} = (1.6 \pm 0.1) \delta A_{\rm n}(T^*_{\rm m}) + (100 \pm 4).
\end{equation}

The following  equation correspond to REL2, obtained in Paper~I:
\begin{equation} 
R_{{\rm M}, n+1} = (1.5 \pm 0.1)  A_{{\rm S}, n} (T^*_{\rm M}) +
 (22 \pm 10).
\end{equation}

It should be noted that the difference in the  temporal dependence  in
the  correlations between the sums of
 the areas of the sunspot groups in the latitude intervals  of
 the northern and  
southern hemispheres is due to (or it implies)  the temporal dependence in the 
 north-south (N-S) asymmetry of the sunspot activity~\citep[for details on 
the N-S asymmetry of solar activity see][Papers~I and II, and references therein]{jg97,hyc09}. 
Although  both the relations REL1 and REL2 are based on almost equal 
high correlations, 
they yielded a substantial different values for the amplitude of
 solar cycle~24, $viz.$ 
 $103\pm10$ and  $74\pm10$, respectively.
  The current trend (not shown in Fig. 1) of 13-month smoothed monthly  
$R_{\rm Z}$ indicates two peaks (Gnevyshev peaks)  for   the current solar 
cycle~24:   one with 
value 66.9 in February, 2012  and another with value 75 in 
October, 2013.  
 Obviously the   prediction 
 based on the REL1 seems to be failed.  The value $74\pm10$ predicted from the 
REL2, i.e., by using Equation~(3),   within its uncertainty limits is 
 close to the observed $R_{\rm M}$.

 In the present analysis we determined the long-term periodicities in 
 the differences (N-S asymmetries) between the sums of the areas of the
 sunspot groups in different $10^\circ$ latitude intervals of the northern and 
southern hemispheres by using  FFT and MEM. 
The time-dependencies in the periodicities  are checked
 by using the  Morlet-wavelet analysis.
 We checked whether REL1 and REL2 
are connected to long-term periodicities in the N-S asymmetry  of sunspot 
activity.
 By using  REL2   it is  
possible to predict only the amplitude of a cycle with a good  
accuracy by about 9-year advance. Therefore,  
 here we have also checked whether it is possible  to predict the 
annual mean  $R_{\rm Z}$,  by  determining 
the cross-correlations  between the 
annual mean $R_{\rm Z}$ and 
the annual  sum of the areas of sunspot groups in
 different $10^\circ$ latitude intervals.
That is, we checked whether it is possible  to predict 
the shape and  length of a solar cycle, and also the epoch and the strength 
its minimum. In addition, by using REL1 
and   REL2 (mainly), 
 we predicted the amplitude of solar cycle~25.

It should be noted here that,  for the first time in the solar cycles history,
 in the case of solar cycle 24 the second peak is larger than the first peak.
REL2 seems to be  related to the Gnevyshev gaps as well as the
 epoch of change  in the sign of global magnetic field.
In the current cycle the polarities of  north-pole magnetic fields are
 already changed. Therefore,
although the epoch of maximum of cycle~24 may be  October 2013, we have made
a tentative  prediction for the
amplitude of cycle~25 by using REL2 with reference to the first peak at
February 2012.

In the next section we will describe  data and method of analysis.
 In Section~3 we will 
present  results, and in Section~4 we will present  conclusions and 
a brief discussion.

\section{Data and method of analysis} 
In  Papers~I and II   we have used the Greenwich sunspot group data 
during  1874\,--\,1976 and the 
 SOON sunspot group data 
during  1977\,--\,2006. Now SOON sunspot group data are  used for seven 
more years, 2007\,--\,2013 (the data are available up to date) 
and are also taken from 
the David Hathaway's  website 
{\tt http://solarscience.msfc.nasa.gov/greenwich.shtml}.  
The data analysis  is carried out by  
  taking all the precautions that were taken  in Papers~I and II.
 We determined the annual variations in the
 sums of the areas of sunspot groups
in different  $10^\circ$ latitude intervals. The calculations were done 
 for each year data separately and also  
 by  binning the data into  
 3-year moving time intervals (3-year MTIs)
  successively shifted by one year during the
 period 1874\,--\,2013, for the sake of better statistics. 
 The time series of  
the  sums of the areas of 
sunspot groups in different  $10^\circ$ latitude intervals are 
cross-correlated with the annual time series of $R_{\rm Z}$. 

The long-term periodicities in the north-south differences (N-S asymmetries)
  of the sums of the areas of 
sunspot groups in different  $10^\circ$ latitude intervals are determined 
from the FFT analysis, the values of the periodicities are determined 
 by using MEM, and the Morlet wavelet analysis is used to find 
the temporal 
dependence in the periodicities.
In the case of long and evenly spaced time series (windows), Fourier transform
 method is a powerful tool, whereas in the  case of short time series 
the Fourier 
components can only determined at harmonics of the fundamental frequency.   
  Besides FFT we have used here  MEM  because it is a  powerful 
spectral  method
 for detecting
the relevant longer periodicities from a short time series~\citep[a brief description about the MEM 
and the wavelet analysis  is given below, also see][]{jj11}.

Essentially MEM involves fitting an autoregressive model 
to the data based on the principle that the resultant spectral 
estimate should be based on all the information in the actual record
 and assume the least possible amount of information 
(hence the name maximum entropy) about the series outside the 
observed record. The condition for this happens to be 
satisfied by the $\mu$th order autoregressive process. 
An important step in this method is the optimum selection of $\mu$,
 which is the number of immediately previous points that have been used
in the calculation of a new point.
If $\mu$ is chosen too low the spectrum is over-smoothed and the high resolution
potential is lost. If $\mu$ is chosen too high, frequency shifting and
spontaneous splitting
 of the spectral peaks occurs,  giving a  false peaky  appearance to the 
spectrum. Objective methods for selecting $\mu$ have been suggested by 
a number of authors~\citep[see][]{barton83}.
The MEM code which we have used here takes the values for $\mu$
 in the range ($\tau$/3, $\tau$/2) \citep{ub75}    or 
$2\tau/\ln(2\tau)$
 \citep{berry78}, where $\tau$ is the  size of a time series.
 We have computed
MEM power spectra choosing  various
values for   $\mu$ in the range ($\tau$/3, $\tau$/2) and  $2\tau/\ln(2\tau)$.
We find that $\mu = {\tau/3}$ is  suitable in the present MEM
analysis,  $i.e.$, in the derived spectra the peaks are considerably  sharp
and  separated.

Wavelet  analysis helps to find simultaneously the periodicities in a
 time series and the temporal dependence in the corresponding amplitudes 
of the periodicities.  
We have used the wavelet  IDL code
provided by Ch. Torrence and G. P. Compo as described in
 \cite{tc98}. A wavelet is a function with zero mean and that is 
localised in both frequency and time. 
The Morlet wavelet is defined as 
$$\psi_0(\eta) = \pi^{-\frac{1}{4}} e^{i\omega_0 \eta} e^{-i\frac{1}{2} \eta^2}\ ,$$ 
\noindent where $\omega_0$ is dimensionless frequency and $\eta$ is 
 dimensionless time.  The Morlet wavelet (with $\omega_0 = 6$) is a good choice, 
because it provides a good balance between time and frequency 
space~\citep{grinsted04}.
 The wavelet is stretched in time by varying its scale ($s$) and
 normalising it to have unit energy. Thus, $\eta = s.t$, where $t$ is time.  
The wavelet scale $s$ is almost identical
to the corresponding Fourier period, i.e.,
 the Morlet wavelet with $\omega_0 = 6$ gives $\lambda = 1.03 s$,
where $\lambda$ is the Fourier period.
Regions where edge effects become important, because of the finite length of the time series,
are labelled as cone of influence (COI). The time series is
padded  with sufficient zeroes to bring the total length of the time series  
up to the next power of two,  limiting the edge effects
and speeding up the
 Fourier transform~\citep[for more details see][]{tc98}).

\section{Results}

 Fig.~2 shows the variations in  the annual sums  of the
 areas of the sunspot groups,  $A_{\rm N} (t)$ and   $A_{\rm S} (t)$, in 
 $0^\circ-10^\circ$ latitude intervals of the northern and
  southern hemispheres  during the period 1875\,--\,2013.
 In the same figure  
 the variations in    
 $A_{\rm N} (T^*_{\rm m})$,  $A_{\rm S} (T^*_{\rm M})$,   
and the annual mean of  
  $R_{\rm Z}$ are also shown. 
Fig.~3 is the same as Fig.~2 but  the variations in  
$A_{\rm N} (t)$ and   $A_{\rm S} (t)$ 
 are determined by binning the sunspot group data  during the period
of 1875\,--\,2013 into  3-year MTIs. In this case
 $t = 1976,\ 1977,\dots,2011, \ 2012$ 
   are the middle years of the 3-year MTIs
1875\,--\,1877, 1876\,--\,1978,\dots,2010\,--\,2012, 2011\,--\,2013,
 respectively.
 In these  figures it is easy to identify the locations of   $T^*_{\rm m}$ 
and  $T^*_{\rm M}$.  
 We use the 3-year MTIs
time series because to avoid the gaps in the time series (in Fig.~2
it  can bee  seen that the values of the sums of the areas of sunspot groups 
are zeros during the minima of some solar cycles).
Fig.~4 shows the variations in 
 the  north-south difference $A_{\rm N} (t) - A_{\rm S} (t)$, where
 $A_{\rm N} (t)$ and $ A_{\rm S} (t)$  are determined from 3-year MTIs 
(which are shown in Fig.~3).
In this figure we have also shown    $\delta A(T^*_{\rm m})$ and
$\delta A(T^*_{\rm M})$. 

As can be seen in Figs.~2 and 3, the epochs of the maxima and the minima
 of $A_{\rm N} (t)$ and $A_{\rm S} (t)$) 
are not exactly same as those of 
$R_{\rm Z}$. In fact, there is a suggestion of a few year phase difference 
between each of these quantities  and $R_{\rm Z}$. The phase difference 
 is determined below. As can be seen in Fig.~4, the N-S asymmetry, i.e. 
$A_{\rm N} (t) - A_{\rm S} (t)$  is considerably vary
 and there is a suggestion for   
the existence of  a few long-term periodicities in it.

 Fig.~5 shows the FFT power spectra of 
 the  N-S asymmetry (N-S difference) in the sums
 of the areas of the sunspot
groups in each different $10^\circ$ latitude intervals. 
The corresponding MEM and wavelet spectra are shown in Figs.~6 and 7, 
respectively.  As can be seen in Fig.~5, in the FFT power  spectrum  
that correspond to  $0^\circ-10^\circ$  latitude interval, 
the peak  at the corresponding  frequency of $\approx$ 8.5-year periodicity 
is significant on slightly 
above 99\% confidence level ($2.8\sigma$). The corresponding  peak is
 highly insignificant in the FFT power spectrum that correspond to       
 $20^\circ-30^\circ$  latitude interval. There is a relatively 
  less significant ($1.85\sigma$)
 peak at the corresponding frequency of   9.7-year periodicity 
in the FFT power
 spectrum that correspond to    $10^\circ-20^\circ$  latitude interval. 
In this spectrum there is   also  a
 peak at the corresponding frequency of  $\approx$15.1-year periodicity, 
 which is  significant on 98\% confidence level.  
In the corresponding  FFT spectra of the $0^\circ -10^\circ$ and 
$10^\circ -20^\circ$ latitude intervals there are peaks at frequency 
correspond to the $\approx$12.4-year  
periodicity,  which are significant on   slightly larger than  
96\% confidence level ($2.1\sigma$\,--\,$2.5\sigma$). 
All the three spectrum show   peaks at the corresponding 
frequency of   60\,--\,100 year periodicity. The  peak at this frequency   
is relatively weak (statistically insignificant) in the case of the
spectrum that correspond to $0^\circ-10^\circ$ latitude interval.  
The corresponding MEM spectra indicate the values of all these periodicities as  66.7-year,  16-year, 12.5-year, and 8.7-year.  
There are  other
relatively short periodicities (7-year, 5.5-year and 4-year), but they
 can be seen
mainly in the MEM spectrum that correspond to  $20^\circ - 30^\circ$ 
latitude interval.
Overall the known $\approx 9$  year periodicity 
   in the N-S asymmetry of the sunspot area ~\citep{hyc09}
 exists mainly in below $20^\circ$ latitudes 
 (it is very weak in 
$20^\circ - 30^\circ$ latitude interval).
 The $\approx$ 12.5-year periodicity, which is  found in N-S asymmetry of
sunspot activity~\citep{carb93,jg97}, 
is much clear  in the  MEM spectrum that correspond to $0^\circ - 10^\circ$ 
latitude  interval. 
 The  44\,--\,55 year cycles can be seen  in both
 $\delta A (t^*_{\rm m})$ and  $\delta A (T^*_{\rm M})$ (see Fig.~4).

As can be seen in Fig.~7, the $\approx 9$-year periodicity exists in all the 
three latitude intervals, but it was mainly exist during the high activity
 period 1940\,--\,1980.  The $\approx$ 12.5-year periodicity exists
  in $0^\circ-10^\circ$ latitude interval mainly during the low activity 
period 1890\,--\,1939. The temporal dependence of 
the  $\approx$ 12.5-year periodicity
is not clear  in the  other two  latitude intervals. 
In the wavelet spectra the $\approx$60-year periodicity is  not well defined,
 i.e.  it is within  the cross-hatched region throughout the period 
1875\,--\,2013. Of course, this periodicity is obviously connected to  the 
two components of the last  Gleissberg cycle: the high activity 9-year 
periodicity  and the 
low activity 12-year periodicity  portions of the 
north-south difference in the sunspot activity 
 at $0^\circ - 10^\circ$ latitude interval. This may indicate that
 the current period could be low activity portion of the ongoing
 Gleissberg cycle.

[In the FFT and MEM spectra (not shown here) of the sums of the areas of 
the sunspot groups 
in  the individual latitude intervals of northern and southern hemispheres, 
 the 11-year periodicity is found dominant in all the three latitude 
intervals of each hemisphere. 
A 8\,--\,9 year periodicity is also seen in the MEM spectra of 
all the latitude intervals of southern 
hemisphere, and  is not seen in the spectrum 
that  correspond to   
 any of the three latitude intervals of northern hemisphere.        
In the wavelet spectra (not shown) that  correspond to the individual 
intervals of the northern and southern hemispheres, the 9-year and 12-year 
 periodicities are found to be unresolved.]

We  calculated the cross-correlations between $A_{\rm N} (t)$  and  annual
 $R_{\rm Z} (t)$,
and also between  $A_{\rm S} (t)$  and annual $R_{\rm Z}(t)$  
(in the case of $R_{\rm Z} (t)$,  $t$ represents actual year).
For this purpose also the time series of 3-year MTIs are used. 
  We  determined the cross-correlations,
 besides in  $\pm\ 0^\circ - 10^\circ$
latitude interval, in $\pm\ 10^\circ - 20^\circ$  and
$\pm\ 20^\circ - 30^\circ$
latitude intervals also. The cross correlations   are  shown in Fig.~8.
 As can be seen in this figure, 
the cross-correlation coefficient ($CCF$) varies   
systematically (damped wave like behaviour)
with  {\it Lag} (absolute), implying that the area-sums vary 
systematically similar to $R_{\rm Z}$. 
In the case of
$0^\circ - 10^\circ$  latitude interval,
the peak of $CCF (R_{\rm Z},\ A_{\rm S})$  at $Lag = -9$ is
   large (with value 0.79)
 and statically high significant as that of the one close to $Lag = +1$
(with value 0.83).   The peak of $CCF (R_{\rm Z}, A_{\rm N})$
 is much smaller (0.66)  at $Lag = -9$,  and  
 slightly
larger (0.83) at $Lag = +1$,   than the corresponding peaks of
$CCF (R_{\rm Z}, A_{\rm S})$.
In the  cases of $10^\circ - 20^\circ$ and $20^\circ - 30^\circ$
 latitude intervals, $CCF (R_{\rm Z}, A_{\rm S})$ and
 $CCF (R_{\rm Z}, A_{\rm N})$ have dominant  peaks
at $Lag = 0$ and $Lag = -1$, receptively and the peaks at $Lag = -9$
are  considerably smaller than the corresponding one of
$CCF (R_{\rm Z}, A_{\rm S})$ in
$0^\circ - 10^\circ$  latitude interval.
The differences between  the peaks of  $CCF (R_{\rm Z}, A_{\rm N})$ and
 $CCF (R_{\rm Z}, A_{\rm S})$
are larger in the case of $0^\circ - 10^\circ$ latitude intervals than
corresponding differences
in the other two latitude intervals, particularly for  
negative values of $Lag$.

A peak of {\it CCF} at a negative value of $Lag$ implies that
 the corresponding  $A_{\rm N}$ or
 $A_{\rm S}$ leads  $R_{\rm Z}$. That is,  in the case of
 $CCF (R_{\rm Z}, A_{\rm S})$ of
$0^\circ - 10^\circ$ latitude interval  the presence
of a dominant peak at $Lag = -9$ implies
 $A_{\rm S}$  leads $R_{\rm Z}$ by about nine years.
This is  consistent with
the high correlation between $A_{{\rm S}, n} (T^*_{\rm M})$ and 
$R_{{\rm M}, n+1}$ ($cf.$ REL2).

  Fig.~9 shows the
scatter plot of $A_{\rm S} (t)$ of the sunspot groups in $0^\circ-10^\circ$
latitude interval of the southern hemisphere versus  $R_{\rm Z} (t+9)$.
In this  figure the continuous curve represents the following linear
  relationship between $A_{\rm S} (t)$ and $R_{\rm Z} (t+9)$:

\begin{equation}
 R_{\rm Z} (t+9) = (0.47 \pm 0.03) A_S (t) + (11 \pm 3) .  
\end{equation}

\noindent The corresponding values of the correlation 
coefficient ($r =  0.85$) and
 the slope are statistically high significant (derived from 128 data 
points,  from the Student's t-test   $P  < 0.01$).

Fig.~10 shows the plots of the   simulated (by using Equation~(4))
 and the observed values of the yearly mean $R_{\rm Z}$ versus time (year).
In this figure it can be seen that the shape of the simulated cycle~24
is  almost same as the pattern of temporal behaviour of
 $A_{\rm S} (t)$
 of the sunspot groups in $0^\circ-10^\circ$
latitude interval of the southern hemisphere during cycle~23 (see Fig.~3).
 The simulated value  $120 \pm 24$ for  $R_{\rm M}$ of cycle~24 
 is almost equal to  the observed as well as the simulated values of 
 $R_{\rm M}$ of cycle~23, a contradiction to
all the corresponding predictions for $R_{\rm M}$ of cycle~24 
 in Papers~I and II. 
 The simulated amplitudes of some small/large cycles  are considerably
 larger/smaller than
 the corresponding observed
amplitudes, causing the  
uncertainty
of the simulated  value is high  ($\sigma = 24$), making  
the reliability of this simulated   $R_{\rm M}$ of cycle~24  
 is lower than the previous predictions that were made
 by using REL2   (or  even that was made by using REL1).
 As per the current level of sunspot activity it seems to be 
almost certain that the simulated value is much higher than the 
real observed amplitude of cycle~24.
In fact, in Paper~II  we have noticed that a higher 
value of predicted $R_{\rm M}$ correspond to 
 a lower value of $r$. 
 The lengths of solar cycles considerably very.
In addition, the wavelet analysis suggests that the long-term periodicities 
in $A_{\rm S}(t)$ are time dependent. 
Consequently, for prediction purpose it is not possible to use 
directly the continuous time series of $A_{\rm S} (t)$ in equal time intervals. 
That is,  it is necessary to take the variations in 
the lengths of solar 
cycle into account.
It should be noted here that even in the case of the prediction of
  $R_{\rm M}$ of a cycle 
by using the polar fields at minimum of the cycle could more uncertain and 
 fail if the polar fields were used early before the start of the 
cycle~\citep{sgaard05}. That is, in our method also 
  it is necessary to find the exact epoch in which  the sum of the areas of 
 the sunspot groups highly correlated with $R_{\rm M}$ of next cycle.    
For the same reason we have  used 
the method described in Papers~I and II and found  REL1 and REL2.   

 In Fig.~10 there is
 an indication that  2011 is the epoch of maximum of cycle~24. 
 There is a  suggestion that this cycle  also may be somewhat long  and
 the next
minimum may  be also 
weak similar to  the last (between cycles~23 and 24) very weak minimum. 
There is also a suggestion that a secondary peak will occur in the year 
 2014. 
 The rise to maximum may be  also
 steep in  cycle~24 as that of cycle~23. Interestingly, within the  year 2011
 a very large peak exists in the monthly 
variation 
in $R_{\rm Z}$ (the corresponding figure  is not shown here).  
Some recent analysis suggested that in the northern hemisphere the maximum is
 already occurred  in the year 2011, but not in the southern hemisphere
 \citep{gopal12}.     
Since we have correlated the 3-year MTI's  sum of the
 areas of sunspot groups with the annual 
mean $R_Z$ (mean of the monthly mean values of $R_{\rm Z}$), one can expect the uncertainties up to 12 months 
in the simulated epochs of maxima and minima. 
However,  overall we can draw 
a conclusion that  the  Eq.~(4) is not much useful for 
 prediction purpose.
(There is a considerable spread in Fig.~9.
 The error in the sum of the areas of sunspot groups   is an
 essential part of the 
present analysis, but that error is difficult to generate here because 
the values of the errors in the areas of sunspot groups are not
 available/known.)

 REL2  is essentially a special case of the 
 $A_{\rm S} (t) - R_{\rm Z}$ 
relationship  (cf., conclusion~2 and Eq.~(4)) correspond to 
 the peak of {\it CCF}  at $Lag = -9$ year. 
 The peak of {\it CCF} at  $Lag = 1\ {\rm to}\ 2$ years   suggest
 the existence of a relation between  $R_{\rm M}$  and  
$A_{\rm S} (T^*_{\rm M})$ of 
a  cycle, but we find a very weak correlation ($r = 0.34$) between these
 parameters of a cycle. That is, $R_{\rm Z}$ of a year may have an effect
 on $A_{\rm S}$ of the following year, but may not be beyond one year. 
Therefore, it seems $R_{\rm M}$ of 
cycle does not have  a strong affect on  $A_{\rm S}(T^*_{\rm M})$ of the 
same cycle. 
$A_{\rm S} (T^*_{\rm M})$ may represent the proxy of  new magnetic flux  
 or largely
 the magnetic flux that transported from the northern hemisphere do to the 
equatorial cross of the flux at $T^*_{\rm M}$. It should be noted here,      
 $T^*_{\rm M}$ is close to the epoch of global change in the sign of the
 solar magnetic field. That is, 
overall it seems  REL2  related to the
global change in the sign of the solar magnetic field.
The $T^*_{\rm M}$ of cycle $n$ and $T_{\rm M}$ of cycle $n+1$ are at the
  beginning and at about one year before the ending  of a  half solar
magnetic cycle, respectively. It is not clear whether the REL2 related to the
  approximate 9-year periodicity in the north-south difference of the
 sum of the areas of sunspot groups in the $0^\circ-10^\circ$ latitude 
intervals of the northern and southern hemispheres, because the  wavelet 
analyses suggest that the approximate 9-year periodicity does not exist
continuously  throughout 1875\,--\,2013.

\section{Prediction for the amplitude of solar cycle~25}
As already mentioned in section~1, although   the values of 
of the corresponding correlation coefficients of both the
 relations REL1 and REL2 are  almost equal and   
high, 
they yielded a substantial different values for the amplitude of
 solar cycle~24, $viz.$  $100\pm10$ and  $74\pm10$, respectively.
 A plausible reason for this could be due to the missing
of some normalisation factor between these relationships. 
On the other hand, 
 the REL2 is better defined than that of REL1, in the sense that the 
intercepts
 of the corresponding linear equations have a small ($22\pm10$) and a
 large  ($74\pm 10$) values, respectively. That is,   
 one-to-one correspondence between $R_{\rm M}$ and 
$A_{\rm S}(T^*_{\rm M})$ is much stronger than that 
  between  $R_{\rm M}$   and    $A_{\rm N}(T^*_{\rm m})$. 
Observationally,  the uncertainty in  $A_{\rm N}(T^*_{\rm m})$ 
is more  than that in $A_{\rm S}(T^*_{\rm M})$ because  the data  
near the maximum of 
a solar cycle is
  well measured(/defined) than that at the minimum.
 By solving  Eq.~(1) that 
correspond to REL1 and Eq.~(3) that correspond to REL2, we get a new 
equation 
 which is free of constant term. From this way in Paper-I 
  it was obtained  
 $57\pm13$  for $R_{\rm M}$ of cycle~24. Similarly, if we
solve  Eqs.~(2)  and (3), we get 

\begin{equation}
R_{{\rm M}, n+1} \approx 1.98 \times  A_{{\rm S}, n} (T^*_{\rm M}) -  0.46 
\times \delta A_n (T^*_{\rm m}) .  
\end{equation}

 By using Eq.~(5) we get 
$66\pm13$ for $R_{\rm M}$ of cycle~24 which is almost equal to the value of
  the first  peak of cycle~24.   
As per the current level of sunspot activity, it is almost certain that 
this,  and the value $74\pm10$ that was predicted by using REL2 alone,   
will be very close to the reality.

The   value 
of about 100 found from REL1 for amplitude of cycle~24 
  seems to be certainly failed. 
However,   by using REL1 it seems still 
 possible to make a  prediction 
 qualitatively ($i.e.$, it seems to be possible to find whether the next
 cycle is stronger or weaker than the immediate previous one)  by 13-year
 advance (that was the major conclusion about the values  that were
 obtained by using 
 REL1 in Papers I and II). For this,  one can also utilise the following relationship:
$A_{{\rm S},n}(T^*_{\rm M}) \approx 1.5 A_{{\rm N},n}(T^*_{\rm m})+ 50$.
That is, first obtain the approximate value of 
$A_{{\rm S},n}(T^*_{\rm M})$ by using this relationship and then by
 substituting that  in Eq.~(3). 
  (In Paper~II   some other simple  additions of the values of the
 parameters of  REL1 and REL2   were carried out 
and obtained few other values, obviously they are  incorrect though 
each one of them is less than the amplitude of cycle~23. Even these earlier 
predictions indicated that certainly cycle~24 will be weaker than cycle~23).

The maximum (in  $R_{\rm Z}$) of  a  solar cycle
 is not smooth. Two or more peaks can be identified
during the solar maxima and are called Gnevyshev peaks, because
 this splitting of activity was
identified for the first time by
Gnevyshev~(\citeyear{gne67}, \citeyear{gne77}).
The time interval between these peaks, where the level of activity is
relatively low,
 is known as the Gnevyshev gap \citep[see the review by][]{sto03}.
Incidentally, the epochs of $T^*_{\rm M}$  seem to be close to
the Gnevyshev gaps  and also close to the time when  the sign of the
   magnetic field is changing in the
 northern and southern hemispheres.
 At these occasions there
could be interactions of the plasma flows  of opposite polarities  causing
 cancellation of
magnetic flux. (At these occasions of a cycle downward flow may also
taking place, which may provide required resource for next cycle.)
 Overall, it seems  REL2  related to the
global change in the sign of the solar magnetic field. In fact,
the $T^*_{\rm M}$ of cycle $n$ and $T_{\rm M}$ of cycle $n+1$ are at the
  beginning and near the ending  of a same half solar
magnetic cycle, respectively.
\cite{kgb11} suggested that the double peaks and the Gnevyshev gap
 in sunspot maximum can be explained as the sunspot maximum is the
superposition of the two surges of toroidal magnetic field generated
 by two parts of the poloidal field. One part generated at the surface is
advected by the meridional circulation all the way to the poles and
 another part is diffused directly to the tachocline at mid-latitudes.
One can also think that the cause of the Gnevyshev peaks/gaps may be due to
interaction  of at least
two magnetic waves   whose   amplitudes and phases are  different. Therefore,
  Eq.~(5),  which is  the linear combination of Eqs.~(2) and (3),
may be having  the aforementioned  physical background rather than mere
mathematical. 

Here  we have used  SOON sunspot data for seven more years 
and found  that the latest   $T_{\rm m}^*$
 and $T_{\rm M}^*$  
(with reference to the first peak of $R_{\rm Z}$ at 2012.170)  
are 2007.55\,--\,2011.05  and 2013.17\,--\,2013.92,  
and the values of the corresponding 
$A_{\rm N} (T^*_{\rm m})$ and $A_{\rm S} (T^*_{\rm M})$  are 0.48  
and 18.86, respectively.

In  principle, 
according to REL1, $i.e.$ by using Eqs.~(1) and (2), it should be 
possible to make a  prediction for 
$R_{\rm M}$ of solar cycle~25. By  using 
Eq.~(2)  we obtained $84\pm 10$ for $R_{\rm M}$ of cycle~25, 
which is much lower than the
corresponding predicted value ($100\pm10$) for $R_{\rm M}$ of cycle~24, but 
  $R_{\rm M} \approx 75$ of cycle~24 is 
 much less than 84.  Therefore,   this result only indicates that 
the cycle~25 could be relatively weaker (not represents the 
amplitude of cycle~24), suggesting that the REL1 (Eqs.~(1) and (2)) 
can be used to make a qualitative prediction for the amplitude of a cycle. 
That is,  Eqs.~(1) and (2) can be used to  
 know  whether the amplitude of next cycle is less or greater than 
that of previous one, by about 13 year advance.  
 
On the other hand by using REL2,  $i.e.$ by using Eq.~(3), 
 we can make a reasonably accurate prediction.
  By using using Eq.~(3)  
we get $50 \pm 10$ for
 $R_{\rm M}$ of cycle 25. By using Eq.~(5) we get  $42 \pm 13$ 
(Note: in this case  the latest value of $\delta A_n (T_{\rm m}^*) = -9.77$).

We would like to  make the following
clarifications about our methods of analysis:
 Fig.~11 shows the changes  in the  correlations
 as the changes in $T^*_{\rm m}$
 and $T^*_{\rm M}$.
 Here we have used 1.0 year  (instead of 0.05 year used in Papers~I and II)
   increase/decrease  in each successive iteration.
 The end points of the time intervals  (represented by the
interval numbers)
are away from
  $T_{\rm m}$ and $T_{\rm M}$ as follows: Intervals
 1, 2, 3, 4,\dots,7
are  at (-2, -1) (-2, 0),(-2, 1),(-2, 2),\dots,(-2, 5) years away,
respectively;  the end points of the intervals 8, 9,\dots,13  are at (-1, 0)
(-1, 1),\dots,(-1, 5) years away, respectively;  similarly, the end points
of  the    intervals  19, 20,\dots,22  are at (1, 2), (1, 3),\dots,(1, 5)
years away, respectively;
 and those of the intervals  26 and 27  are at (3, 4)
and (3, 5) years away, respectively.  In the intervals 4 and 19
 the corresponding correlation coefficients have large values,
  0.85 and 0.91, respectively.
From  $T_{\rm m}$ and $T_{\rm M}$
  the end points of the intervals 4 and 19  are at  (-2, 2) and
 (1, 2) years away, respectively, and obviously
$T^*_{\rm m}$  and  $T^*_{\rm M}$ are within these
intervals, respectively.
  The
increase/decrease time 0.05 year that was used in Papers I and II  for
 accurately determining $T^*_{\rm m}$ and $T^*_{\rm M}$
 may be small,
but  $T^*_{\rm m}$, and even $T^*_{\rm M}$,
  are relatively long and the corresponding correlations are
 still high   with plus or minus of  a few times of 0.05 year
 from these intervals. That is, there is a considerable consistency
in the data sampling.

 In Fig.~11  (which demonstrates the methodology used in Papers~I and II)
there are steep changes in the correlation coefficient
 correspond to   the data in $0^\circ - 10^\circ$ latitude interval of
 the southern hemisphere, and  the variation in the correlation coefficient
correspond to  the data  in $0^\circ - 10^\circ$ latitude interval of
 the  northern hemisphere is  somewhat gradual.
In Fig.~2 one can find that
 the aforementioned steep/gradual changes  seen in Fig.~11 is a
property of the solar
cycle variations in the sum of the areas of the sunspot groups
 in   $0^\circ - 10^\circ$ latitude interval of
 the southern/northern
hemisphere. That is,  $T^*_{\rm M}$
is close to the
epoch of maximum (peak) of the solar cycle variation in the sum of the areas
 of the sunspot groups
  in   $0^\circ - 10^\circ$ latitude interval of the southern hemisphere,
where obviously the  variation is  steep. The   $T^*_{\rm m}$
is close to the epoch
of minimum of the solar cycle variation in the sum of the areas of
sunspot groups
  in   $0^\circ - 10^\circ$ latitude interval of the northern hemisphere, where
the variation
is  somewhat gradual. These results  strongly indicate
that  $T^*_{\rm m}$  and  $T^*_{\rm M}$  are reasonably well defined and
  hence both the    relations REL1 and REL2 are reliable.

\section{Conclusions and discussion}
 We  analysed the combined  Greenwich and SOON  sunspot group data during 
the period  1874\,--\,2013 
and found the following results:
\begin{enumerate}
\item In  the difference between the sums of the  areas of the sunspot groups
 in  $0^\circ-10^\circ$ latitude intervals of northern and southern 
hemispheres,   there exist $\approx$ 9-year  periodicity  during the 
high activity  period 1940\,--\,1980 and   $\approx$ 12-year periodicity  
during the low activity period 1890\,--\,1939.
\item There exists a
 high correlation (85\% from 128 data points) between the sum of the
 areas of the sunspot groups in $0^\circ - 10^\circ$ latitude interval of  the
southern hemisphere during  a $Q$th year   and  ($R_{\rm Z}$) of ($Q + 9$)th
 year.
\item By using the above 
relationship (conclusion~2)  it is not possible to make an accurate prediction
 for the amplitude of a solar cycle. 
\item Because of the lengths of solar cycles considerably very and 
the long-term periodicities
 in $A_{\rm S}(t)$ are time dependent (conclusion~1), it is necessary to 
find the exact epoch in which  the sum of the areas of 
 the sunspot groups highly correlated with $R_{\rm M}$ of next cycle.    
Thus, REL2 is a special case of the above relationship (conclusion~2).  
 By using REL2 (Eq.~(3)) it is possible for predicting the amplitude of 
a solar cycle with a  good accuracy.   
\item We predict $50 \pm 10$ for the amplitude of the next cycle~25, 
 which is about
31\% lower than the amplitude of cycle~24 (in fact, it seems to be  
lowest compare with the amplitudes of cycles~1\,--\,24.).  
\end{enumerate}

According to  Gnevyshev and Ohl rule~\citep{go48} or G-O-rule
an odd numbered cycle is always stronger than its preceding even
 numbered cycle. 
By using the G-O rule it will not be
 possible to
predict the amplitude of solar
 cycle~25  when the amplitude of solar cycle~24 is known,
 unless the epochs of the violations of the G-O rule is known in
well advance.
Using the relations REL1 and REL2
  it is possible to
 predict $R_{\rm M}$ of any cycle, $i.e.$ an even--numbered solar cycle as well as
 an odd--numbered solar cycle (no
need to  know in advance the epochs of violations of the G-O rule).
The violation of
 the G-O rule is followed by a few (at least one) relatively weak solar cycles
(see Fig.~1).
For example,  solar cycle pair (4, 5) was followed by weak solar cycles~6
 and 7,  solar cycle pair (8, 9) was also followed a relatively week cycle 10.
 Thus,  one or two weak solar cycles which follow solar cycle pair (22, 23)
 may be also relatively weak.
The low values predicted above for solar cycles~24 and 25 are
 consistent with this long-term property of solar activity. The predicted
amplitudes  of cycles~24 and 25  suggest that the G-O rule will
be violated by this cycle pair (24, 25) also.  That is, it seems for the
 first time  violation of G-O rule  takes place in two consecutive pairs of
 even- and odd-numbered  solar cycles, (22, 23) and (24, 25),
since G-O rule was discovered.

 The long term trend, in fact,
 extrapolation of the cosine curve shown in Fig.~5 of \cite{jbu05}
 from the epoch of
 cycle~23,  suggests that solar
 cycles~24 and 25 will be relatively weak cycles.
  The  above predicted amplitudes
 of  solar cycles~24 and 25  are
consistent with this result.
Thus, our predictions are also consistent with the speculation that
 the minimum of the current
Gleissberg cycle will  take place  in cycle~25 \citep{jbu05}.
The current long-term minimum
 (shall include cycles~24 and 25) is at the
end of the current Gleissberg cycle (preceding minimum of the next
Gleissberg cycle) and it may be similar to the recent Modern Minimum
 (or even to the Dalton Minimum).

Since our results/predictions are based on the north-south
asymmetries in the sums of the
areas of the sunspot groups in the heliographic latitudes that are
 close to the  equator, cancellation/enhancement of magnetic flux due
 to  equatorial  crossings of magnetic flux (caused by the solar meridional
 flows)  may have a major rule in the  mechanism behind our
methodology/results.

\vspace{0.3cm}
 \noindent{\large \bf Acknowledgments}

\vspace{0.3cm}
{The author thanks the anonymous referee for useful comments and suggestions.
 Wavelet software was provided by Ch. Torrence and G.P. Compo,
and is available
at {\tt URL: http//poas.colorado.edu/reserch/wavelets/}. The MEM FORTRAN code
was provided to us by Dr. A.V. Raveendran.}

{}

\newpage
\begin{figure}
\centerline{\includegraphics[width=12.0cm]{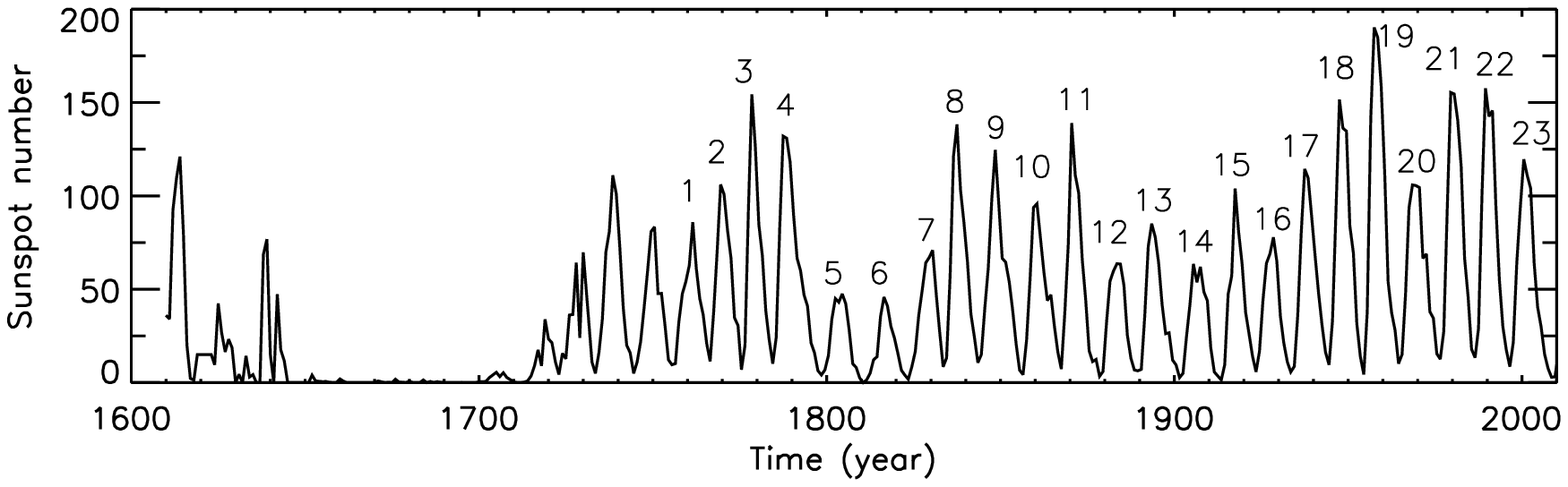}}
\caption{Plots of annual mean international sunspot number ($R_{\rm Z}$) against time. Near the peaks of
the cycles corresponding Waldmeier solar cycle numbers are shown.}
\end{figure}

\begin{figure}
\centerline{\includegraphics[width=11cm]{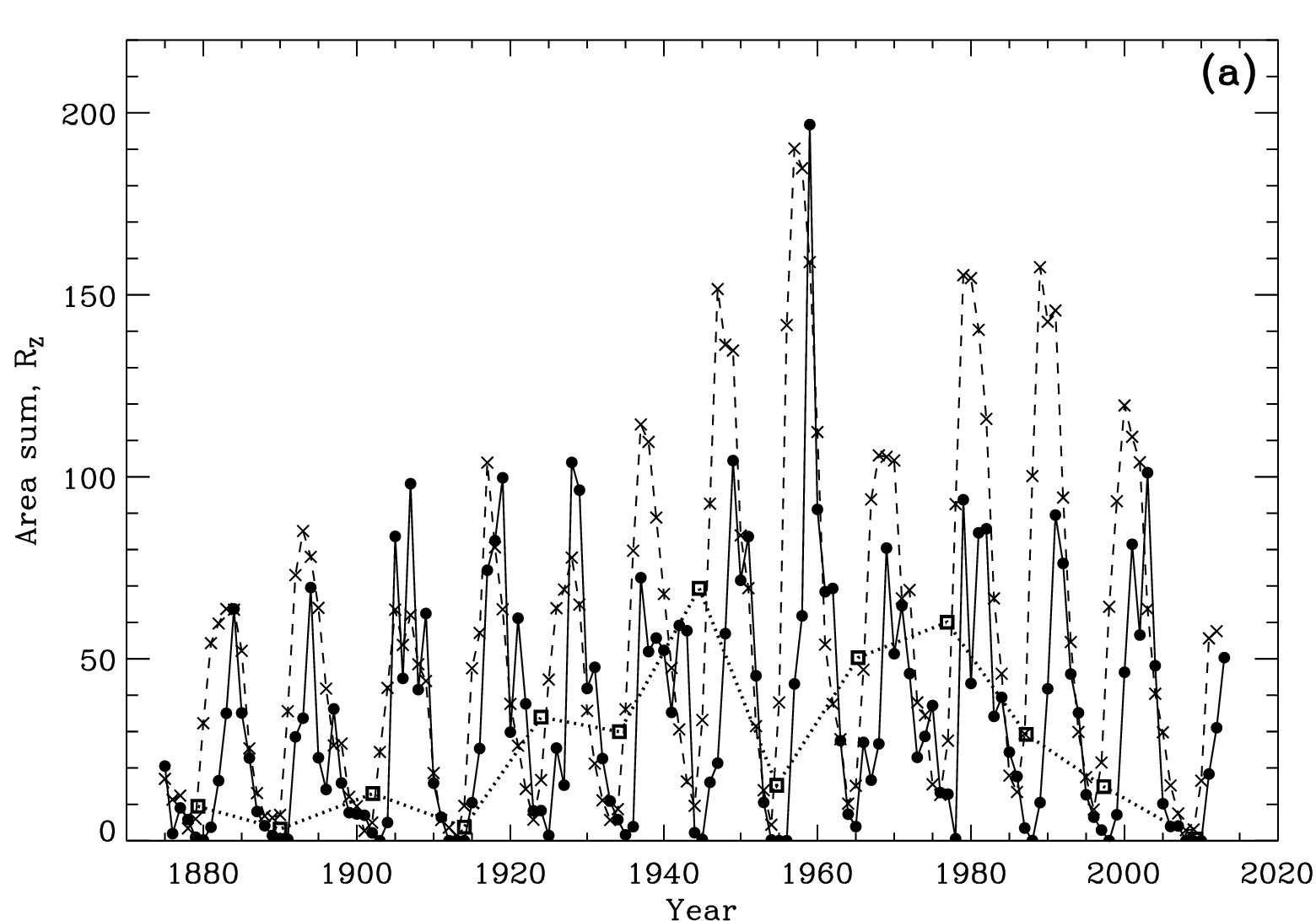}}
\centerline{\includegraphics[width=11cm]{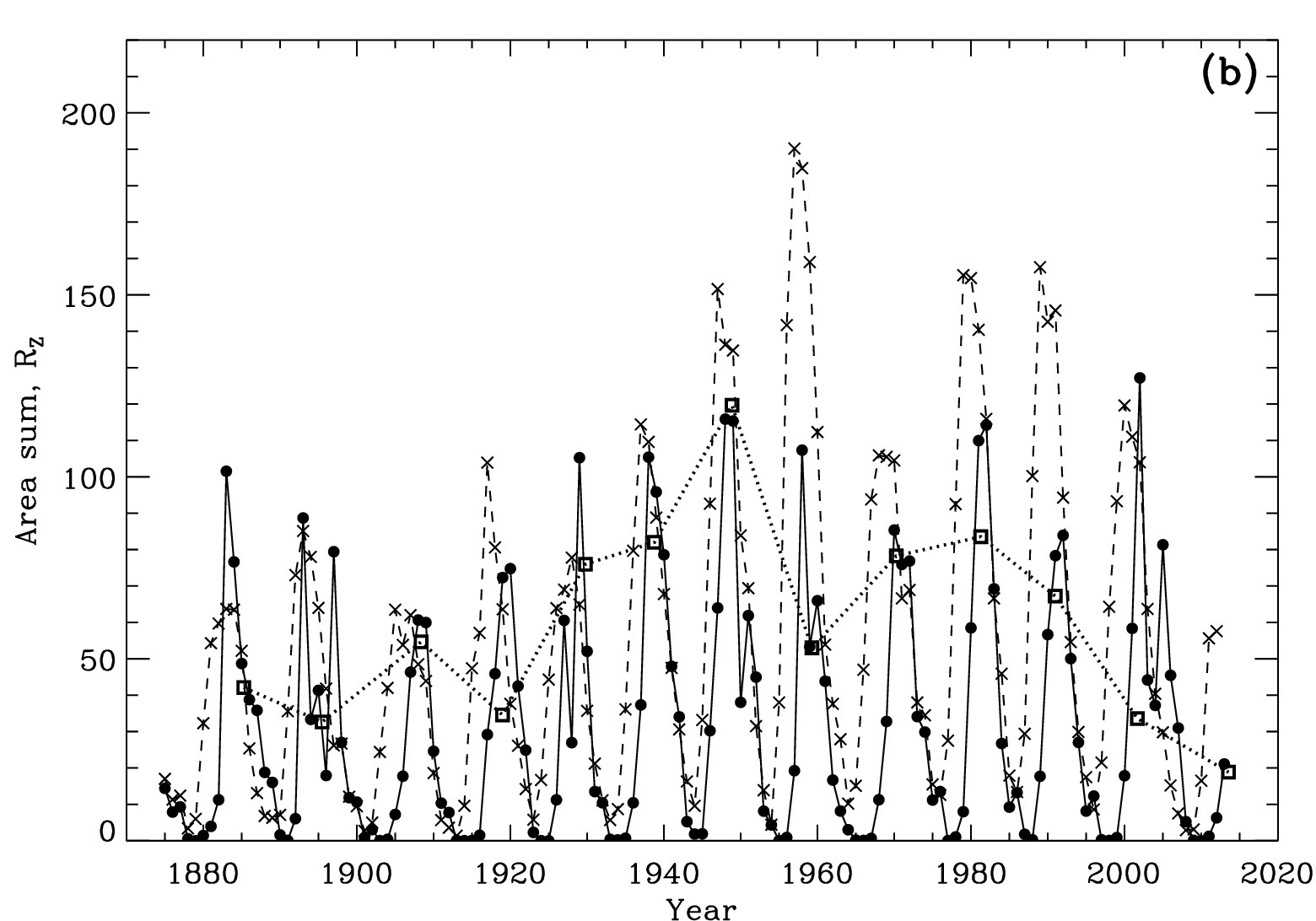}}
\caption{Variations in  the annual  sums (divided by 1000) of the areas
 of the sunspot groups (filled circle solid-curve), $A_{\rm N} (t)$
 and $A_{\rm S} (t)$,   in 
 $0^\circ-10^\circ$ latitude intervals of the northern 
hemisphere (upper panel)  
and the southern hemisphere (lower panel)
   during the period 1875\,--\,2013. 
 The dotted-curves (with squares) represent the plots of   
 $A_{\rm N} (T^*_{\rm m})$ (in upper panel) and $A_{\rm S} (T^*_{\rm M})$ 
(in lower panel)
 versus the middle epochs of   $T^*_{\rm m}$ and $T^*_{\rm M}$,
 respectively.  In both the lower and upper panels the dashed-curve 
(with crosses) represents the annual variation of $R_{\rm Z}$.}
\end{figure}

\begin{figure}
\centerline{\includegraphics[width=11cm]{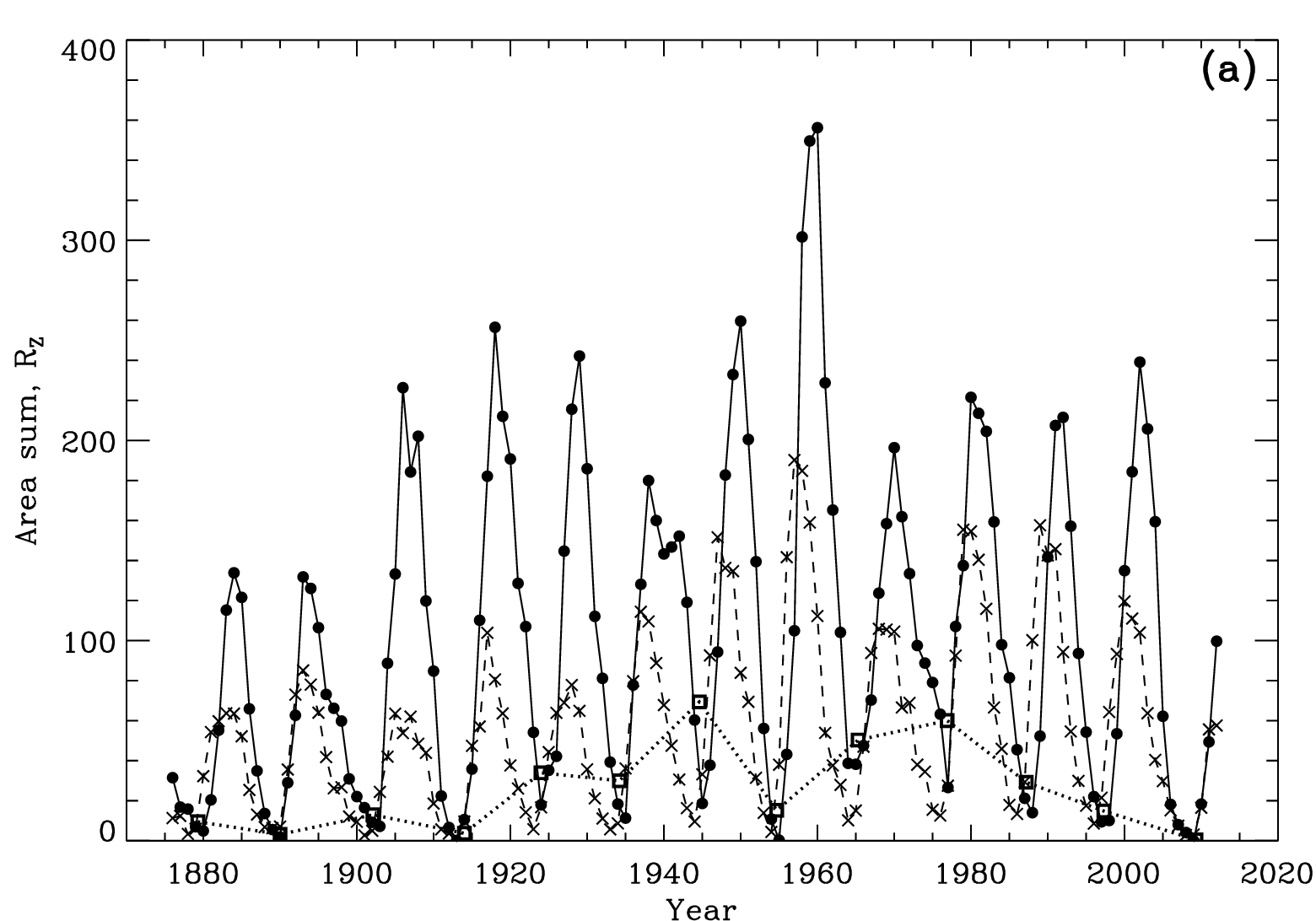}}
\centerline{\includegraphics[width=11cm]{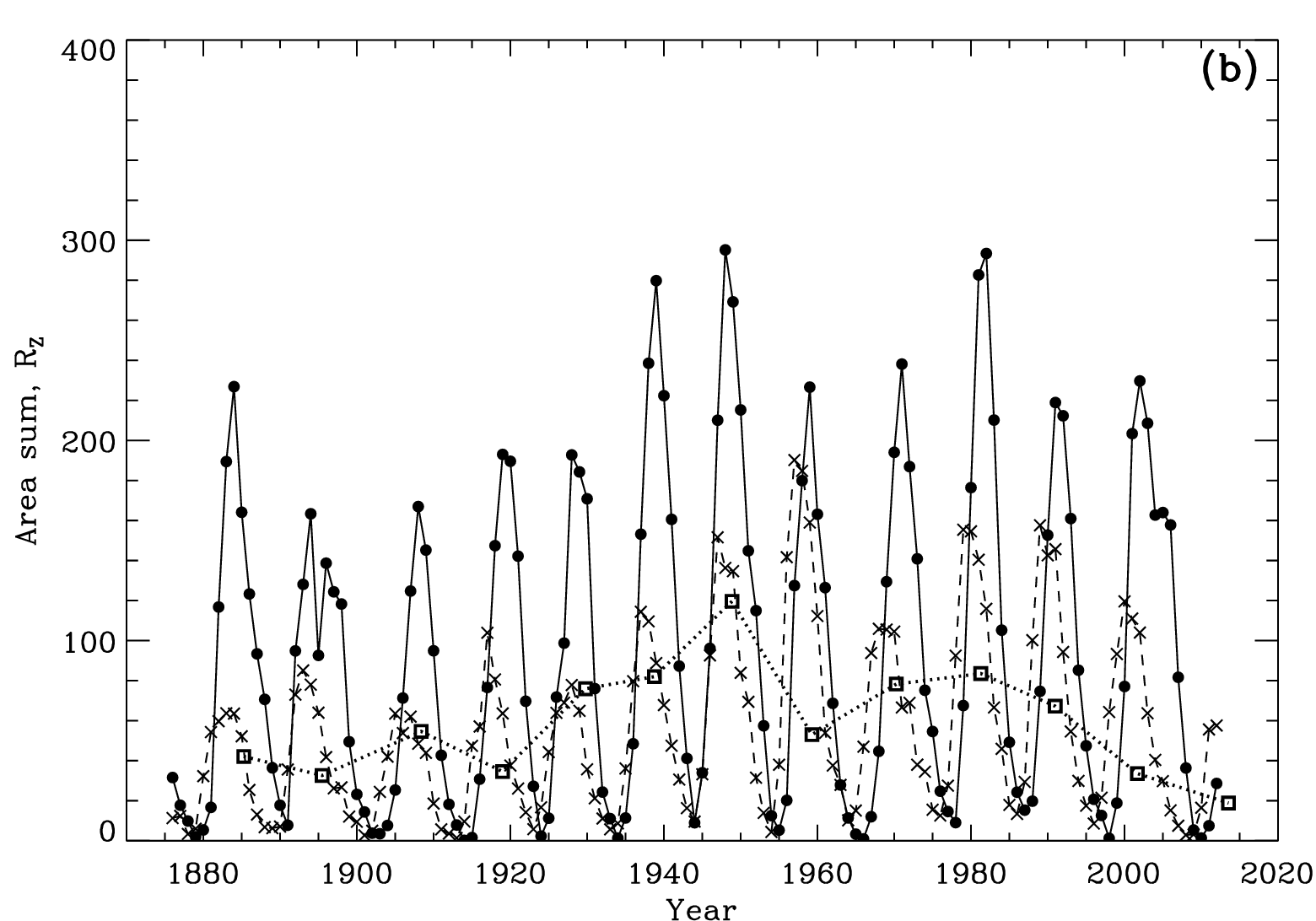}}
\caption{The same as Fig.~2 but the sums of the areas of 
the sunspot groups in
the $0^\circ - 10^\circ$  latitude intervals of the northern
 and southern 
 hemispheres--determined by binning the sunspot group 
 data during the period of
1875\,--\,2013 into  the  3-year MTIs 
1875\,--\,1877, 1876\,--\,1978,...,2010\,--\,2012, 2011\,--\,2013-- 
versus the middle years 1976,\ 1977,\dots,2011,\ 2012 
of the 3-year MTIs.}
\end{figure}

\begin{figure}
\centerline{\includegraphics[width=12.0cm]{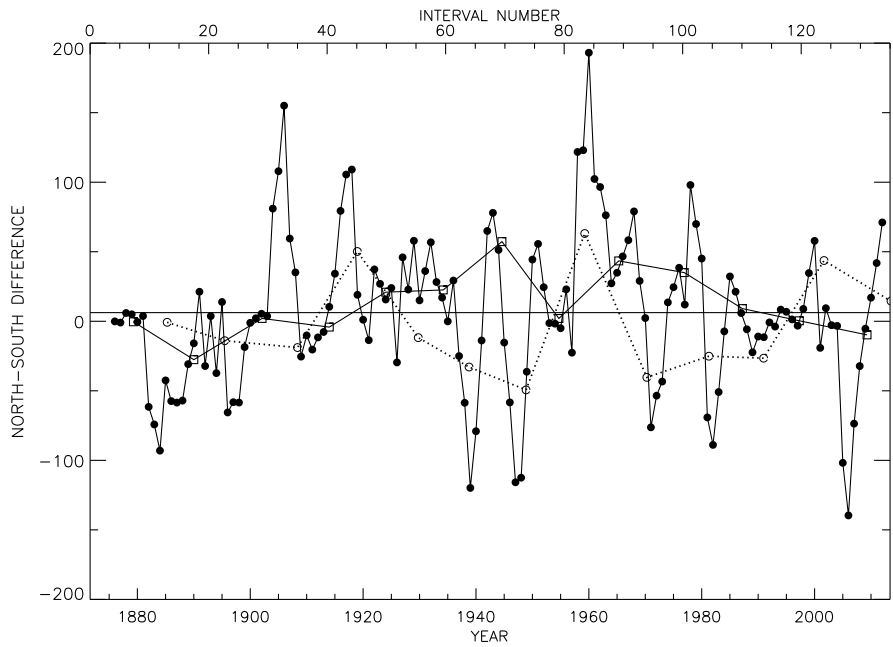}}
\caption{Plot of  the differences between the sums  of the
areas of the sunspot groups in
the $0^\circ - 10^\circ$  latitude intervals of the northern
and southern
 hemispheres  versus middle years of the 3-year MTIs
(closed circle-solid curve).
The square-dashed  and open circle-dotted curves represent the  variations in the
   $\delta A(T^*_{\rm m})$ and
$\delta A(T^*_{\rm M})$, respectively.}
\end{figure}

\begin{figure}
\centerline{\includegraphics[width=9.0cm]{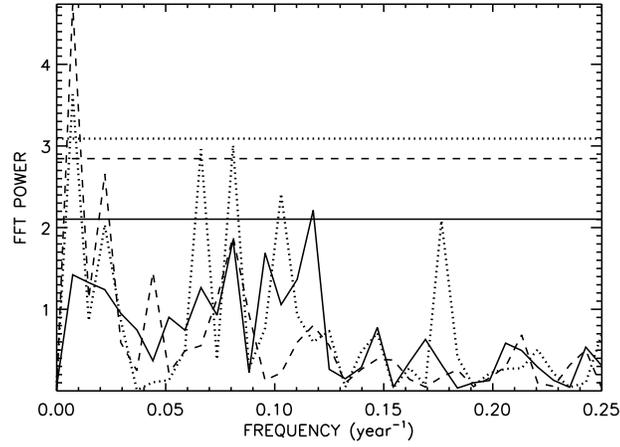}}
\caption{FFT power spectra of the north-south differences in the sums of the
areas of the sunspot groups in $0^\circ - 10^\circ$ (solid curve),
$10^\circ - 20^\circ$ (dotted curve),
 and $20^\circ - 30^\circ$ (dashed curve), determined from 3-year MTIs during
the period 1875\,--\,2013. The horizontal lines represent respective 99\% confidence levels.} 
\end{figure}

\begin{figure}
\centerline{\includegraphics[width=9.0cm]{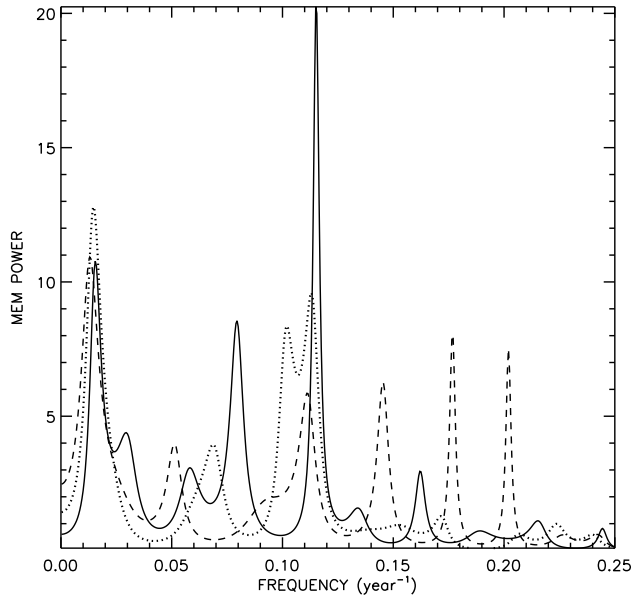}}
\caption{MEM power spectra of the north-south differences in the sums of the
areas of the sunspot groups in $0^\circ - 10^\circ$ (solid curve),
$10^\circ - 20^\circ$ (dotted curve),
 and $20^\circ - 30^\circ$ (dashed curve), determined from 3-year MTIs during
 the period 1875\,--\,2013.}
\end{figure}

\begin{figure}
\centering
\centerline{\includegraphics[width=8.0cm]{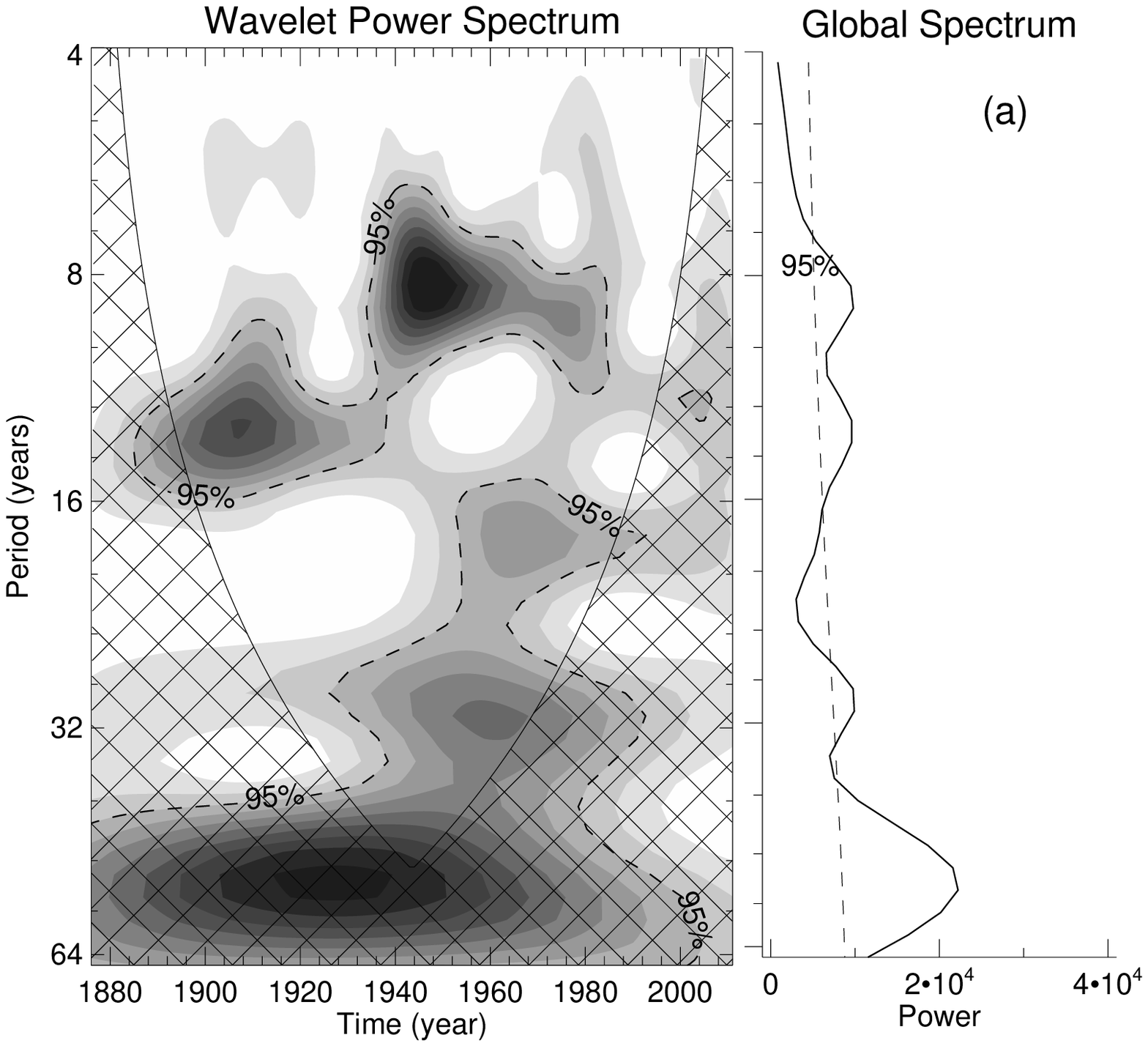}}
\centerline{\includegraphics[width=8.0cm]{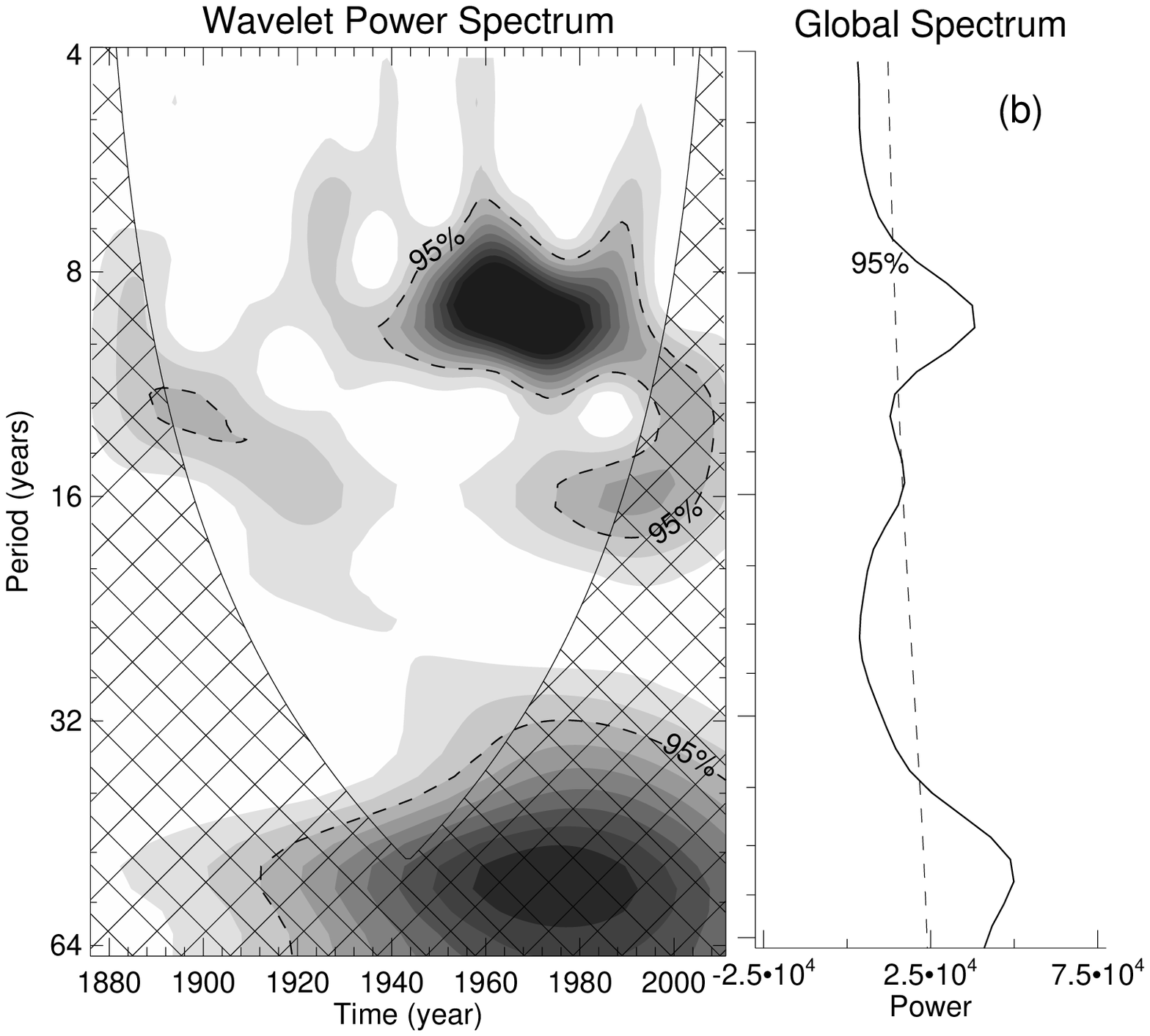}}
\centerline{\includegraphics[width=8.0cm]{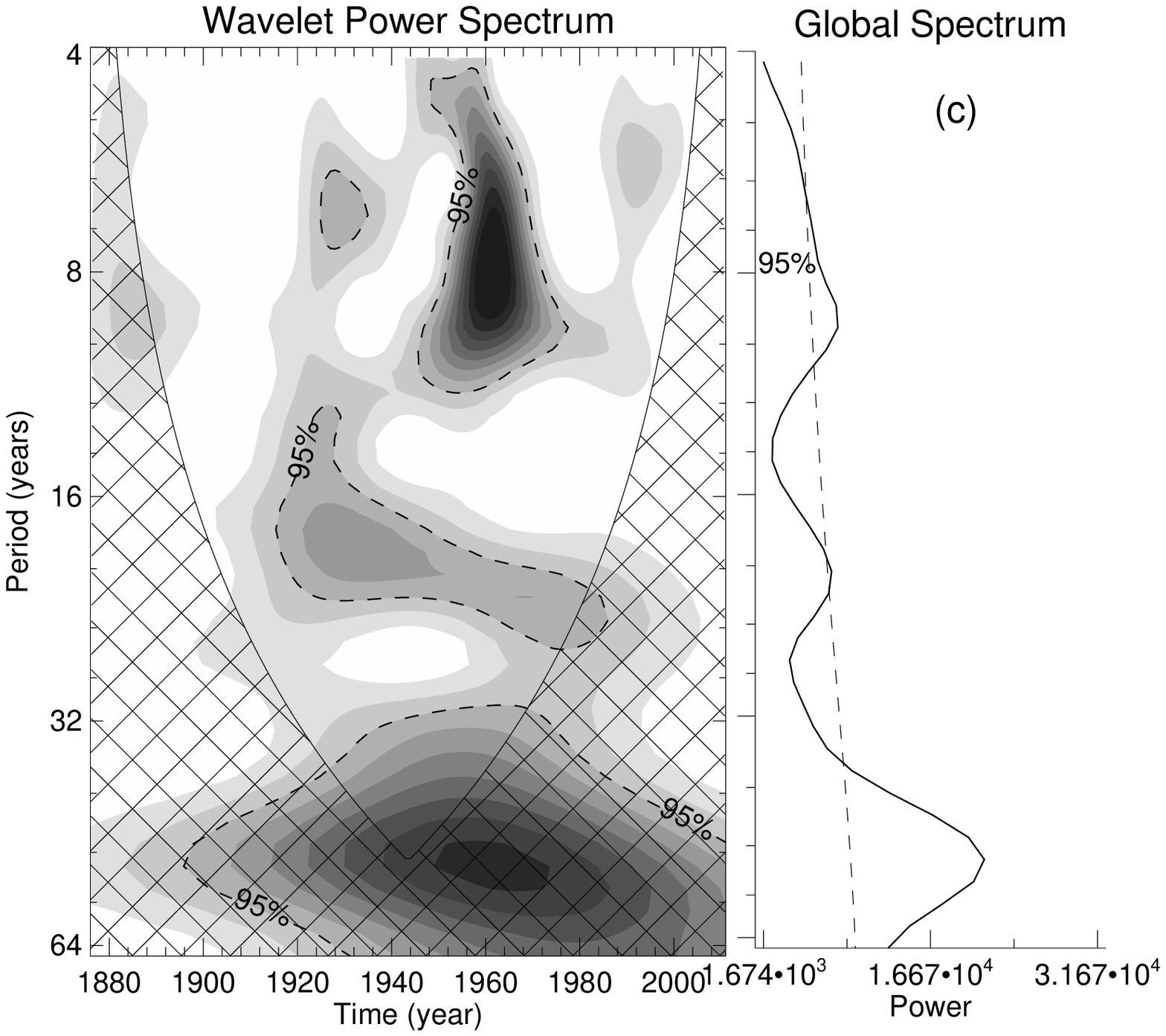}}
\caption{Wavelet power spectra and the global spectra
of the north-south difference in the  sums of 
 the areas of the sunspot groups in
  (a) $0^\circ-10^\circ$,
  (b) $10^\circ-20^\circ$, and
  (c) $20^\circ-30^\circ$ latitude intervals, determined from the data in 
 3-year MTIs during the 
period 1875\,--\,2013. The wavelet spectra are normalised by the variances of
the  corresponding time series.
The shadings are  at the normalised variances of 1.0, 2.0, 3.0, 4.0, 5.0, 6.0,
7.0, 8.0 and 9.0.
 The dashed curves represent the 95\% confidence levels,
 deduced by assuming a white noise process.
The cross-hatched regions indicate the ``cone of
influence", where edge effects become significant (Torrence and Compo 1998).}
\end{figure}

\begin{figure}
\centerline{\includegraphics[width=\textwidth]{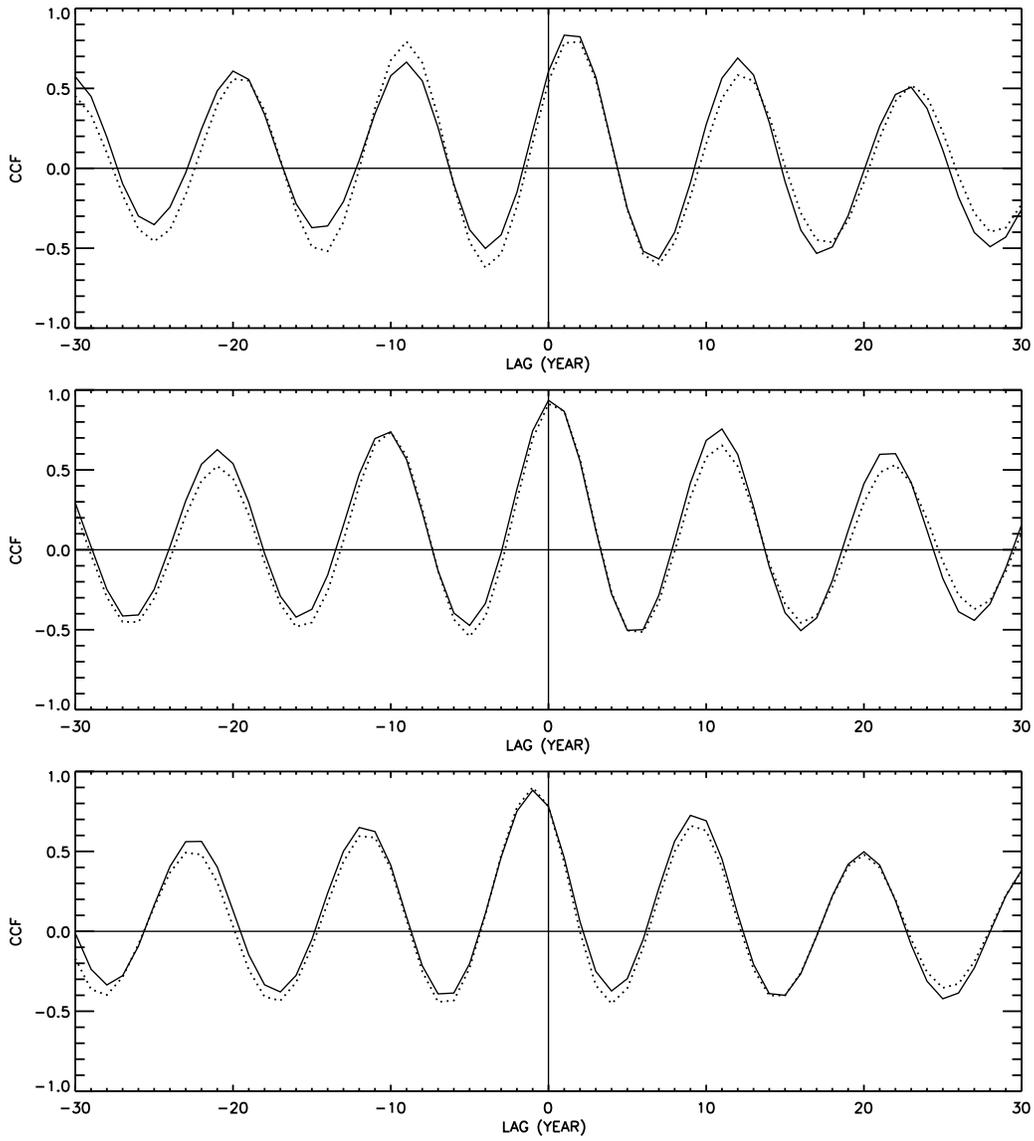}}
\caption{Plots of the cross-correlation coefficients 
 $CCF (R_{\rm Z},A_{\rm N})$ (solid curve) and
 $CCF (R_{\rm Z},A_{\rm S})$
 (dotted curve) versus Lag (year) in $0^\circ - 10^\circ$ (top panel),
$10^\circ - 20^\circ$ (middle panel),
 and $20^\circ - 30^\circ$ (bottom panel).}
\end{figure}

\begin{figure}
\centerline{\includegraphics[width=\textwidth]{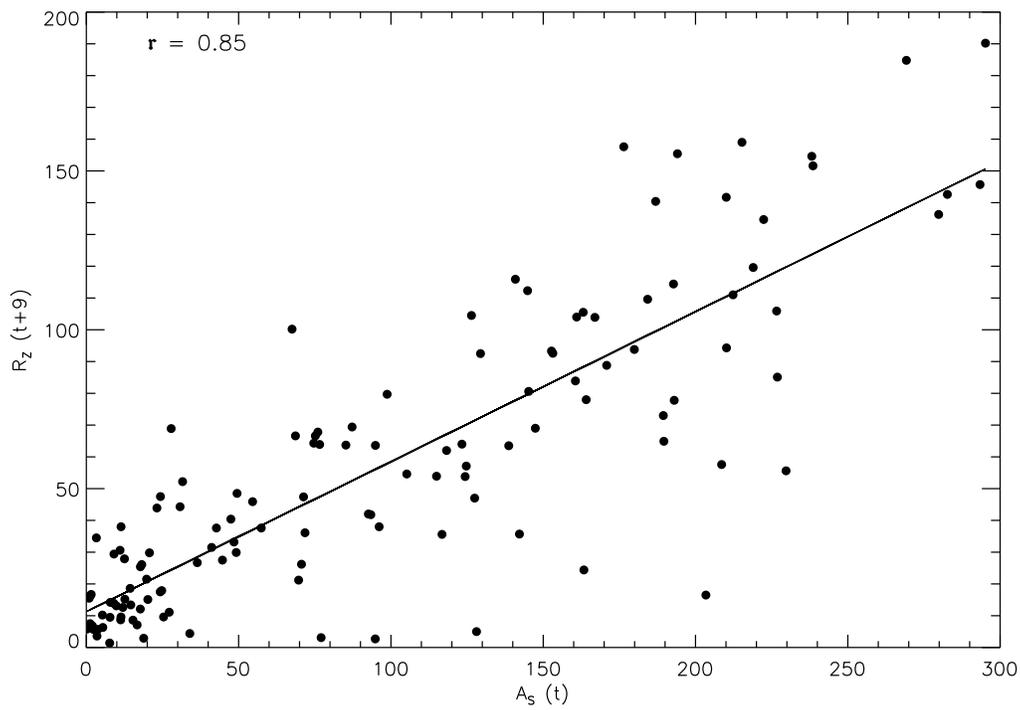}}
\caption{Scatter plot of $A_{\rm S} (t)$
 of the sunspot groups in $0^\circ-10^\circ$
latitude interval of the southern hemisphere versus  $R_{\rm Z} (t+9)$,
 where $t = 1976,\ 1977,\dots,2011,\ 2012$ (in the case of $A_{\rm S} (t)$
  they are the  middle years of the 3-year MTIs
1875\,--\,1877, 1876\,--\,1978,\dots,2010\,--\,2012, 2011\,--\,2013,
 respectively).}
\end{figure}

\begin{figure}
\centerline{\includegraphics[width=\textwidth]{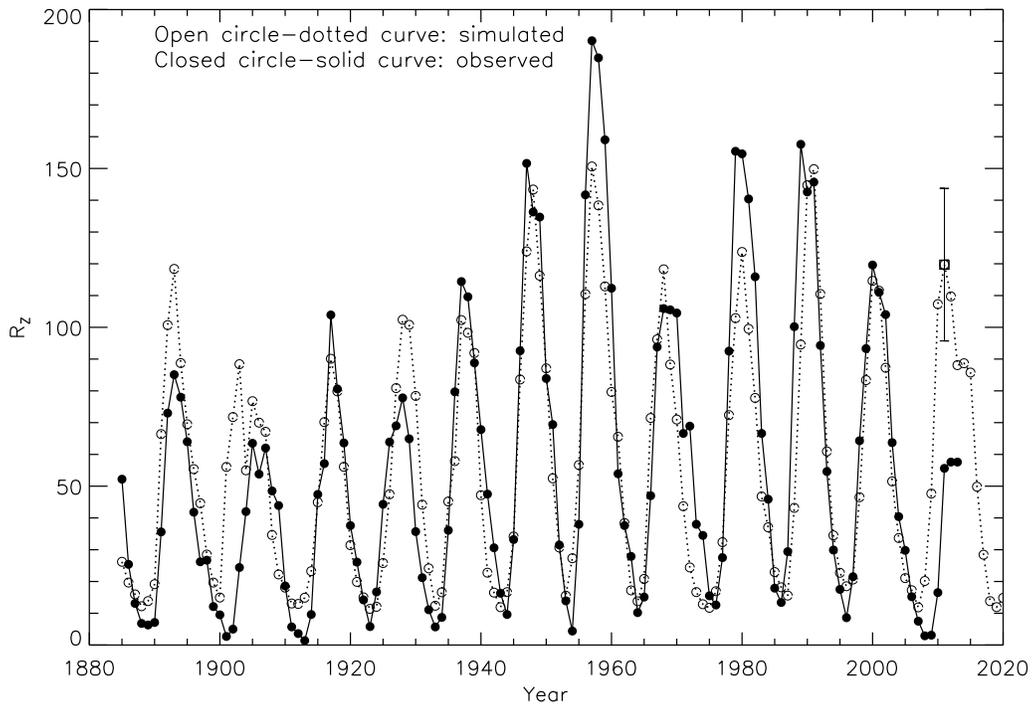}}
\caption{Plot of simulated (by using Eq.~(4)) and observed values 
of $R_{\rm Z}$  versus
 time (year).  The square at year 2011 represents the maximum 
 of the simulated cycle~24  and the error bar represents the standard
 deviation of a simulate value.}
\end{figure}

\begin{figure}
\centerline{\includegraphics[width=\textwidth]{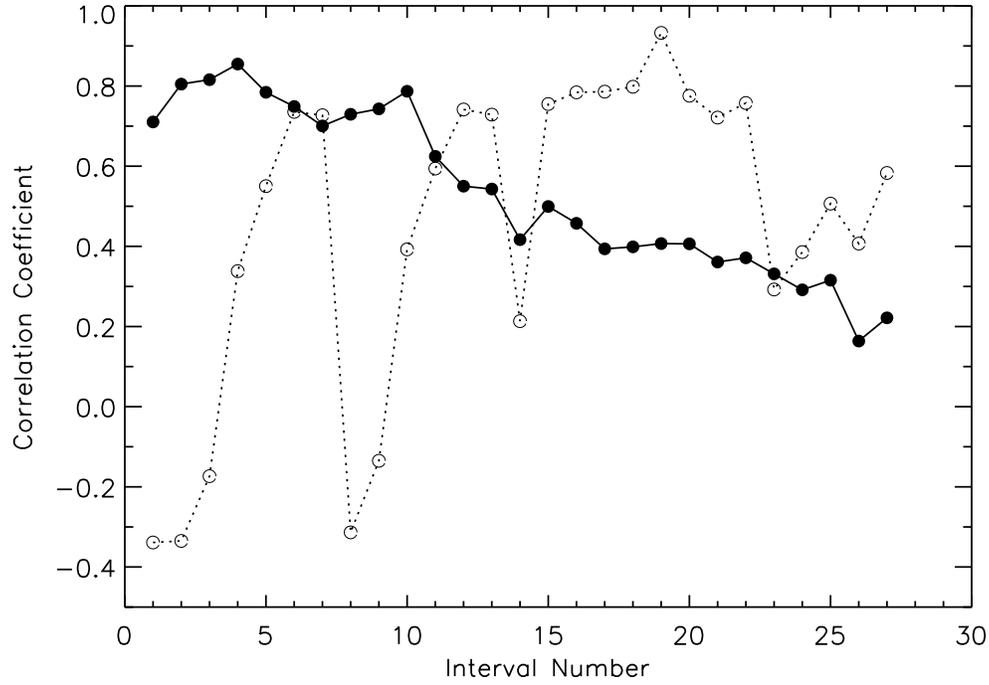}}
\caption{Changes in the  correlations as the changes in $T^*_{\rm m}$
 (solid curve) and in $T^*_{\rm M}$ (dotted curve).
In this figure the end points of the time intervals
(represented by  interval numbers)
are away from
  $T_{\rm m}$ and $T_{\rm M}$ as follows: Intervals
 1, 2, 3, 4,\dots,7
are  at (-2, -1) (-2, 0),(-2, 1),(-2, 2),\dots,(-2, 5) years away,
respectively;  the end points of the intervals 8, 9,\dots,13  are at (-1, 0)
(-1, 1),\dots,(-1, 5) years away, respectively;  similarly, the end points
of  the    intervals  19, 20,\dots,22  are at (1, 2), (1, 3),\dots,(1, 5)
years away, respectively;
 and those of the intervals  26 and 27  are at (3,4)
and (3, 5) years away, respectively.  In the intervals 4 and 19
 the corresponding correlation coefficients have large values,
  0.85 and 0.91, respectively.
From  $T_{\rm m}$ and $T_{\rm M}$
  the end points of the intervals 4 and 19 are at  (-2, 2) and
 (1, 2) years away, respectively, and obviously
$T^*_{\rm m}$  and  $T^*_{\rm M}$ are within these
intervals, respectively.}
\end{figure}
\end{document}